\documentstyle[twoside,fleqn,espcrc2,epsf]{article}
\newcommand{\lsim}{\raisebox{0.3mm}{\em $\, <$} 
\hspace{-3.3mm} \raisebox{-1.8mm}{\em $\sim \,$}}

\begin{document}
\pagestyle{empty}

\title{$\nu_\mu\leftrightarrow\nu_\tau$ vs $\nu_\mu\leftrightarrow\nu_s$
solutions for the atmospheric neutrino problem
\thanks{Talk presented at ``NEUTRINO 98'', Takayama, Japan, 
		June 4-9, 1998.}}

\author{Osamu Yasuda\\
        {\ }\\
        Department of Physics, Tokyo Metropolitan University\\
        1-1 Minami-Osawa Hachioji, Tokyo 192-0397, Japan}

\begin{abstract}
The $\nu_\mu \leftrightarrow\nu_\tau$ and $\nu_\mu
\leftrightarrow\nu_s$ solutions to the atmospheric neutrino problem
are compared with Superkamiokande data.  Both the solutions with a
large mixing angle seem to be consistent with the data.
\end{abstract}

\maketitle


\section{Introduction}

Recent atmospheric neutrino data by Superkamiokande
\cite{sk1,kajita,sk2} provide strong evidence for neutrino
oscillations.  It has been shown \cite{sk1,kajita,sk2} that
atmospheric neutrino data favor $\nu_\mu\leftrightarrow\nu_\tau$
oscillations with maximal mixing, rather than
$\nu_\mu\leftrightarrow\nu_e$.  However, $\nu_\mu$ disappearance alone
does not imply uniquely a $\nu_\mu\leftrightarrow\nu_\tau$ solution
and there is another solution $\nu_\mu\leftrightarrow\nu_s$, where
$\nu_s$ denotes a sterile neutrino.  In this talk some aspects of the
$\nu_\mu\leftrightarrow\nu_\tau$ and $\nu_\mu\leftrightarrow\nu_s$
solutions are discussed.

In the past there has been a prejudice against the
$\nu_\mu\leftrightarrow\nu_s$ solution to the atmospheric neutrino
problem.  The argument \cite{bbn} was based on big bang
nucleosynthesis which gives a condition $\Delta m^2\sin^42\theta\lsim
10^{-4}$eV$^2$ in order for sterile neutrinos not to be in thermal
equilibrium.  However, there was a loophole in this argument.  Foot
and Volkas \cite{fv1} have shown that large lepton asymmetries will
suppress $\nu_s\leftrightarrow\nu_x$ neutrino
oscillations. Interestingly, given certain conditions, the required
lepton asymmetries can actually be created by the oscillations
themselves \cite{fv1}.  So there is no longer any obstruction to
$\nu_\mu\leftrightarrow\nu_s$ as a solution to the atmospheric
neutrino anomaly.

Many models \cite{atmsterile,fv2} have been proposed which predict large
or maximal active-sterile mixing.  Among others, Foot and Volkas have
been obsessed by exact parity symmetric models \cite{fv2} and this was
the main motivation of \cite{fvy} in which
$\nu_\mu\leftrightarrow\nu_s$ was examined in detail by fitting to the
contained events of the Superkamiokande atmospheric neutrino data for
414 days.

\section{Analysis of the Superkamiokande contained events for 414 days}

The survival probability $P(\nu_\alpha\leftrightarrow\nu_\alpha)$ is
obtained by solving the
Schr\"odinger equation for neutrino evolution including matter effects. 
It is given by
\begin{eqnarray}
\displaystyle i {d \over dx} \left( \begin{array}{c} \nu_\mu(x) \\
\nu_{\tau,s}(x)
\end{array} \right)
=M\left( \begin{array}{c} \nu_\mu(x) \\ \nu_{\tau,s}(x)
\end{array} \right),\\
\label{eqn:sch}
M\equiv
U{\rm diag}(0,{\Delta m^2 \over 2E})U^{-1}
+{\rm diag}(0,A_{\tau,s}(x)),\nonumber
\end{eqnarray}
where
\begin{eqnarray}
U\equiv\left( \begin{array}{cc}
\cos\theta&\sin\theta\\
-\sin\theta&\cos\theta\end{array} \right)\nonumber
\end{eqnarray}
is the MNS mixing matrix \cite{mns},
$x$ is the distance traveled, $\Delta m^2$ the difference in squared
masses, $\theta$ the vacuum mixing angle and $\nu_{\mu,\tau,s}(x)$ the
wave-functions of the neutrinos.  The quantities $A_{\tau,s}(x)$
are the effective potential differences generated through the matter
effect \cite{msw}:
\begin{eqnarray}
A_\tau(x) = 0\nonumber
\end{eqnarray}
and, for electrically neutral terrestrial matter \cite{bla}
\begin{eqnarray}
A_s(x) = {1 \over \sqrt{2}} G_F N_n(x),\nonumber
\end{eqnarray}
where $G_F$ is the Fermi constant, $N_n(x)$ is the number density of
neutrons along the path of the neutrino.  It is this matter effect
$A_{\tau,s}$ that make a difference between the
$\nu_\mu\leftrightarrow\nu_\tau$ and $\nu_\mu\leftrightarrow\nu_s$
oscillations.  For antineutrinos the sign of $A_s$ is reversed.

The way to obtain the numbers of events and evaluate $\chi^2$ is described
in \cite{fvy}, where
two quantities have been introduced to perform a $\chi^2$
analysis.  One is the double ratio \cite{kam}
\begin{eqnarray} 
R \equiv \frac{(N_\mu/N_e)|_{\rm osc}}{(N_\mu/N_e)|_{\rm no-osc}}
\nonumber
\end{eqnarray}
where the quantities $N_{e,\mu}$ are the numbers of $e$-like and
$\mu$-like events. The numerator denotes numbers with oscillation
probability obtained by (\ref{eqn:sch}), while the denominator the
numbers expected with oscillations switched off.  The other one is the
quantity on up-down flux asymmetries for $\alpha$-like
($\alpha$=e,$\mu$) events and is
defined by
\begin{eqnarray}
Y_\alpha \equiv
{(N_\alpha(\cos\Theta<-0.2)/N_\alpha(\cos\Theta>0.2))|_{\rm osc}
\over (N_\alpha(\cos\Theta<-0.2)
/N_\alpha(\cos\Theta>0.2))|_{\rm no-osc}},\nonumber
\end{eqnarray}
where $\Theta$ is the zenith angle, $N_\alpha(\cos\Theta<-0.2)$
and $N_\alpha(\cos\Theta>0.2)$
are the number of upward and downward going events, respectively.
$\chi^2$ with the double ratio $R$ is defined by
\begin{eqnarray}
\chi^2_{\rm atm}(R) = \sum_E \left({R^{SK} - R^{th} \over
\delta R^{SK}}\right)^2,\nonumber
\end{eqnarray}
and $\chi^2$ with the up-down asymmetry $Y_\alpha$ is defined by
\begin{eqnarray}
&{\ }&\chi^2_{\rm atm}(Y)\nonumber\\
&=&\sum_E \left[
 \left({Y^{SK}_{\mu} - Y^{th}_{\mu} \over \delta Y^{SK}_{\mu}}\right)^2
+ \left({Y^{SK}_{e} - Y^{th}_{e} \over \delta Y^{SK}_{e}}\right)^2
\right],\nonumber
\end{eqnarray}
where the sum is over the sub-GeV and multi-GeV cases, the measured
Superkamiokande values and errors are denoted by the superscript
``SK'' and the theoretical predictions for the quantities are labeled
by ``th''.

\begin{figure}[p]
\epsfbox[50 50 140 140]{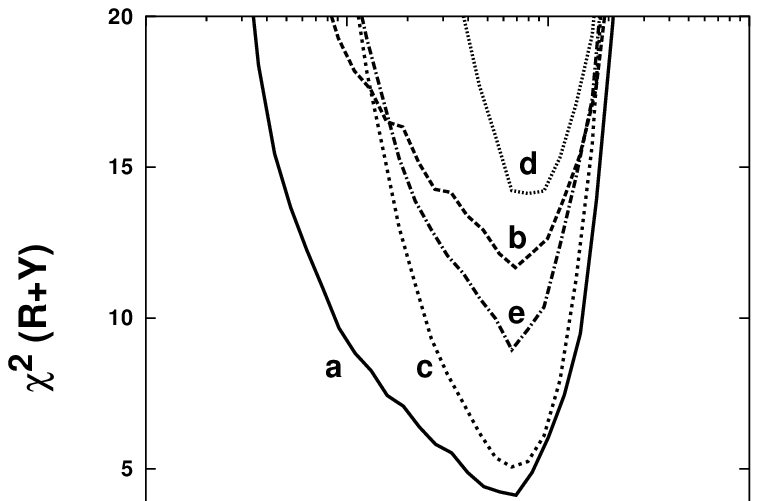}
\vglue 3.4cm
\caption{\label{fig:fig1}$\chi^2$ as a function of $\Delta m^2$.
\vglue 0.01cm
\qquad\qquad(a): $\nu_\mu\leftrightarrow\nu_\tau$, $\sin^22\theta=1$;
\vglue 0.01cm
\qquad\qquad(b): $\nu_\mu\leftrightarrow\nu_\tau$, $\sin^22\theta=0.8$;
\vglue 0.01cm
\qquad\qquad(c): $\nu_\mu\leftrightarrow\nu_s$, $\sin^22\theta=1$;
\vglue 0.01cm
\qquad\qquad(d): $\nu_\mu\leftrightarrow\nu_s$, $\sin^22\theta=0.8$, $\Delta m^2>0$;
\vglue 0.01cm
\qquad \qquad(e): $\nu_\mu\leftrightarrow\nu_s$, $\sin^22\theta=0.8$, $\Delta m^2<0$.}
\end{figure}

\begin{figure}[p]
\epsfbox[50 50 170 170]{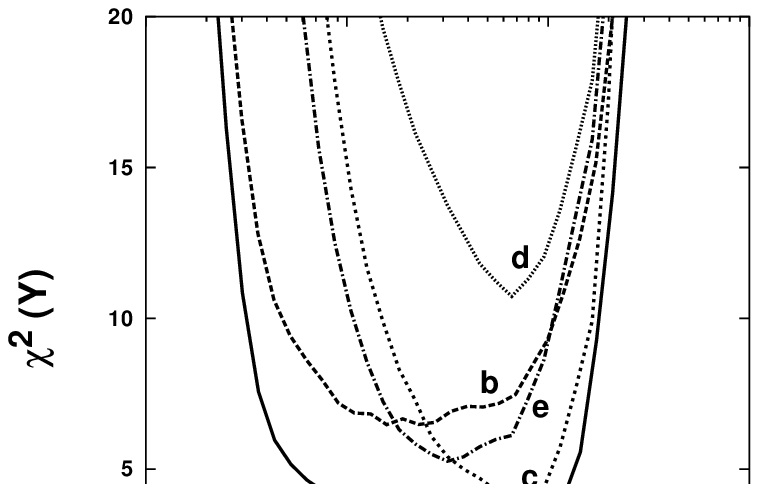}
\vglue 3.1cm
\caption{\label{fig:fig2}The same as Figure 1 but with $R$ data excluded
from the fit.  }
\end{figure}

\begin{figure}[p]
\hbox to \hsize{\hfil\epsfxsize=7cm\epsfbox{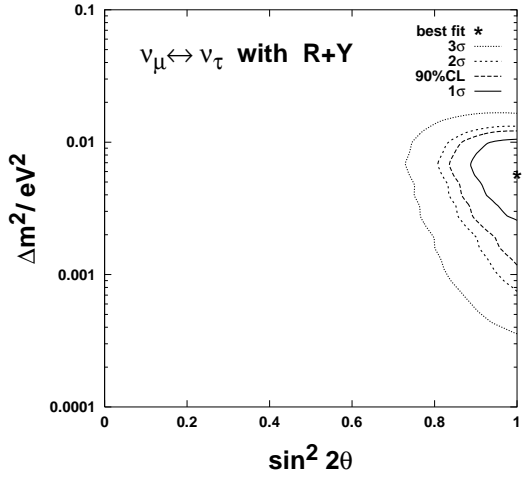}\hfil}
\caption{\label{fig:fig3}The allowed region in the $(\sin^2 2\theta,\
\Delta m^2)$ plane for the $\nu_\mu \leftrightarrow\nu_\tau$ scenario.  }
\end{figure}

\begin{figure}[p]
\hbox to \hsize{\hfil\epsfxsize=7cm\epsfbox{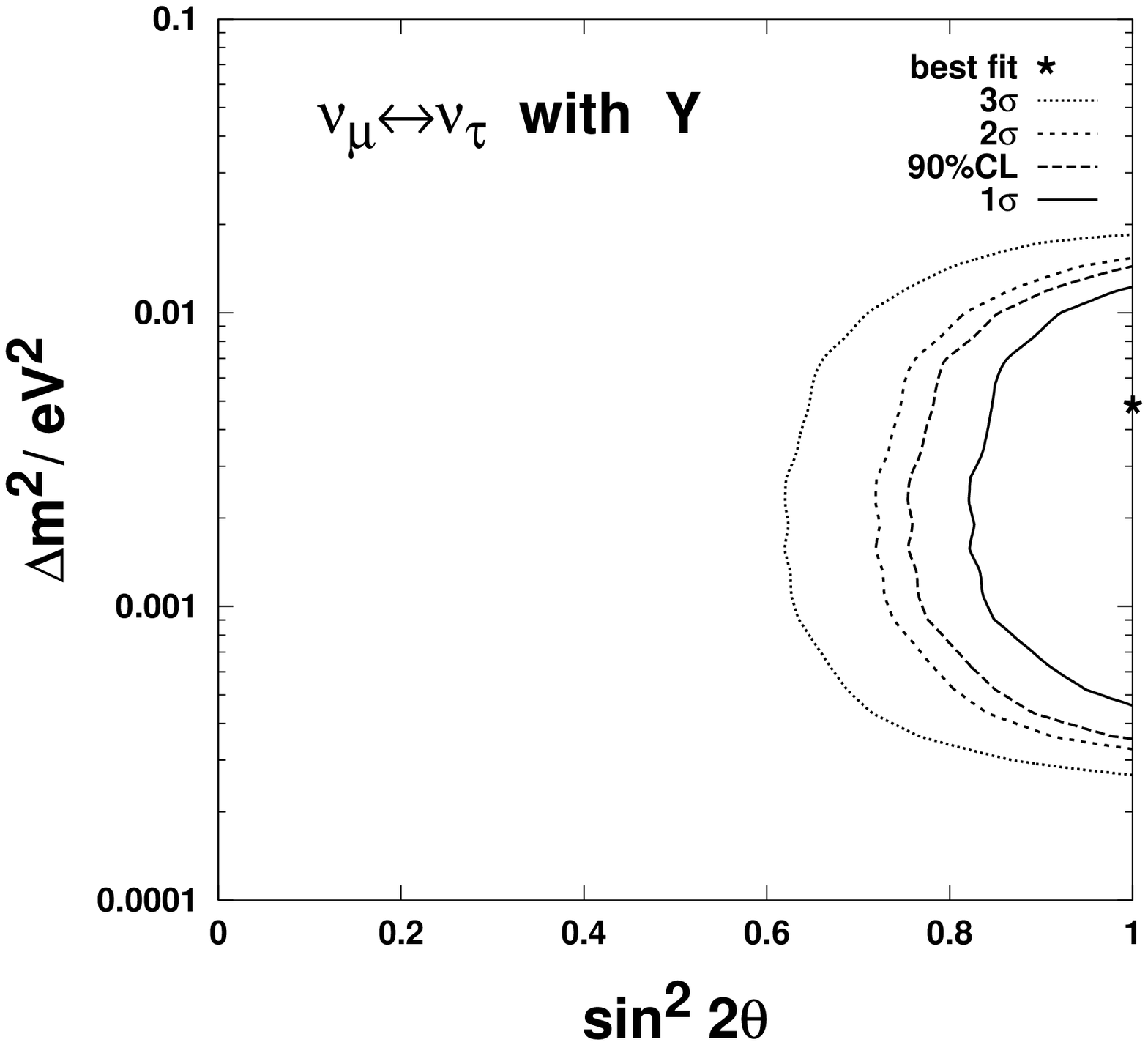}\hfil}
\caption{\label{fig:fig4}The same as Figure 3 but with $R$ data
excluded from the fit.  }
\end{figure}

\begin{figure}[p]
\hbox to \hsize{\hfil\epsfxsize=7cm\epsfbox{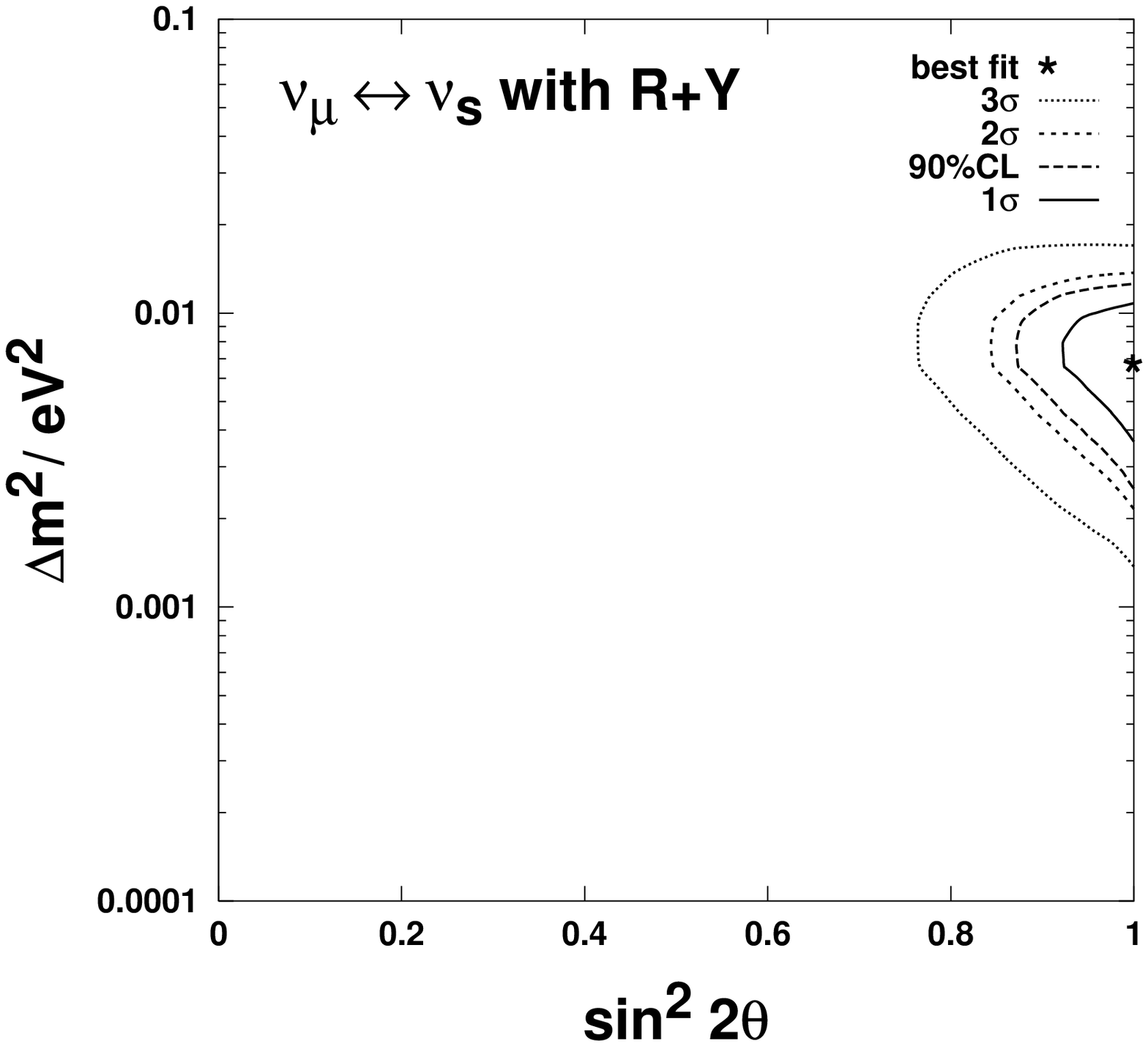}\hfil}
\caption{\label{fig:fig5}The allowed region in the $(\sin^2 2\theta,\
\Delta m^2)$ plane for the $\nu_\mu \leftrightarrow\nu_s$ scenario
with $\Delta m^2 >0$.  }
\end{figure}

\begin{figure}[p]
\hbox to \hsize{\hfil\epsfxsize=7cm\epsfbox{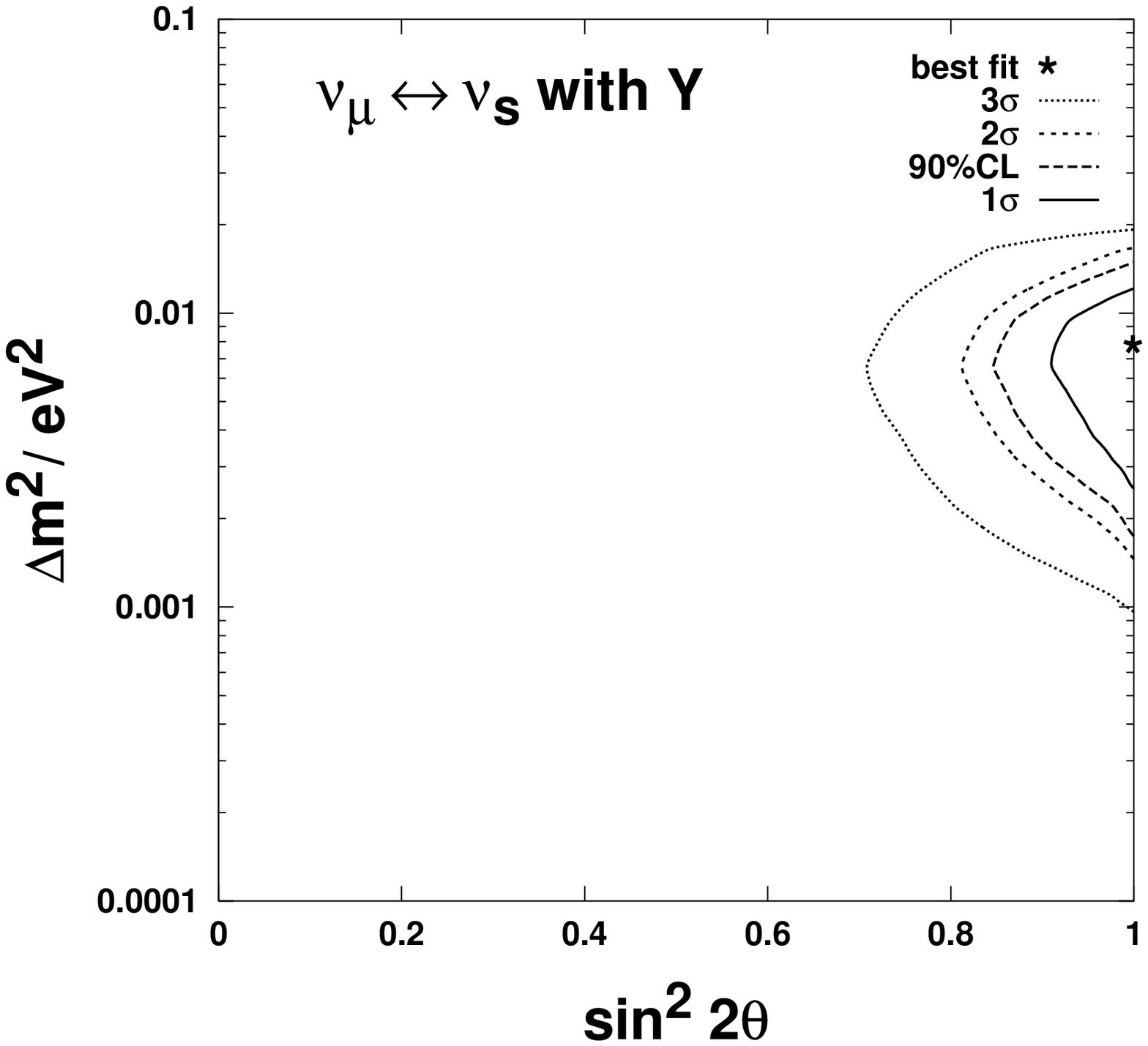}\hfil}
\caption{\label{fig:fig6}The same as Figure 5 but with $R$ data
excluded from the fit. }
\end{figure}

\begin{figure}[p]
\hbox to \hsize{\hfil\epsfxsize=7cm\epsfbox{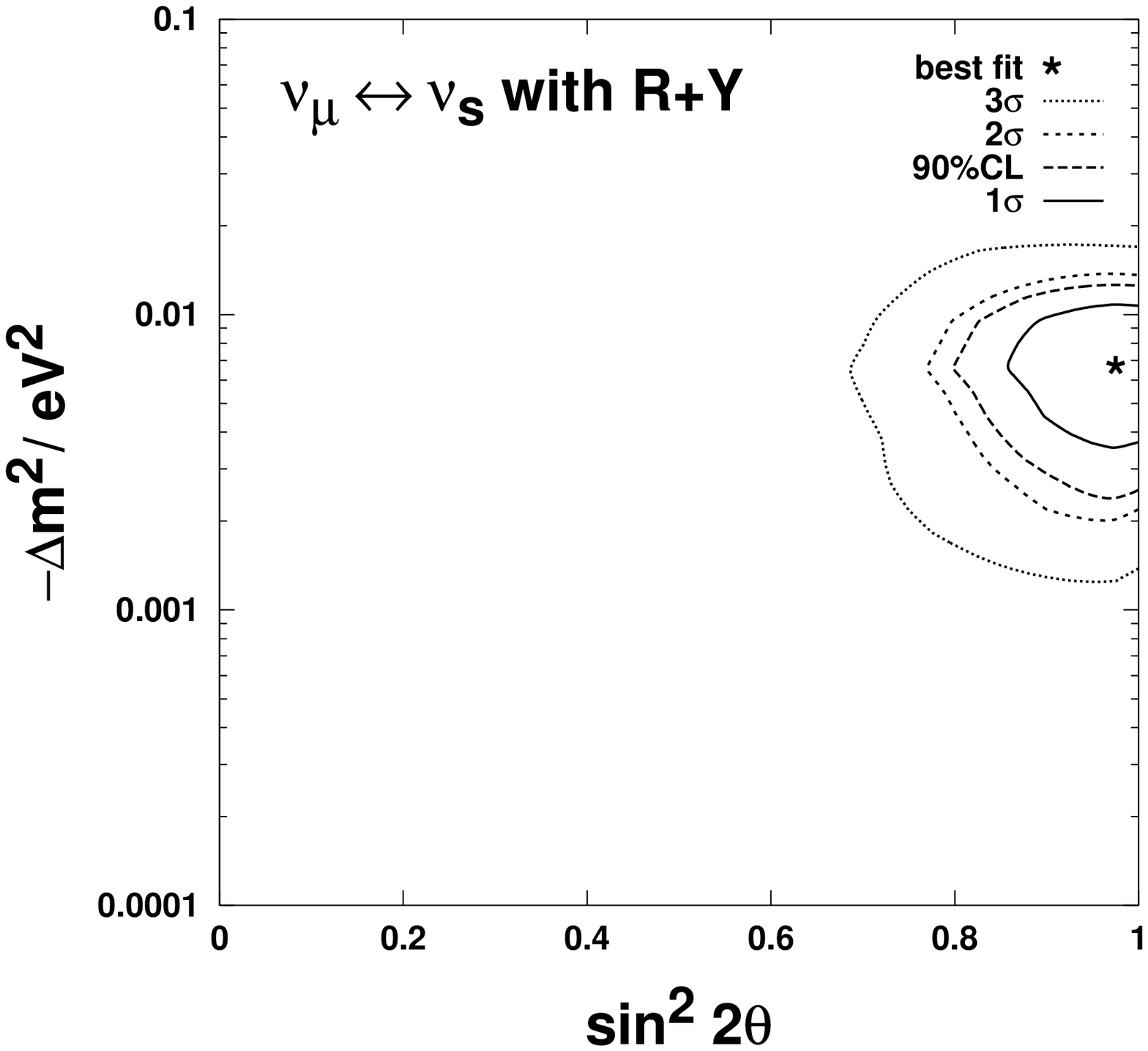}\hfil}
\caption{\label{fig:fig7}The allowed region in the $(\sin^2 2\theta,\
\Delta m^2)$ plane for the $\nu_\mu \leftrightarrow\nu_s$ scenario
with $\Delta m^2 <0$. 
}
\end{figure}

\begin{figure}[p]
\hbox to \hsize{\hfil\epsfxsize=7cm\epsfbox{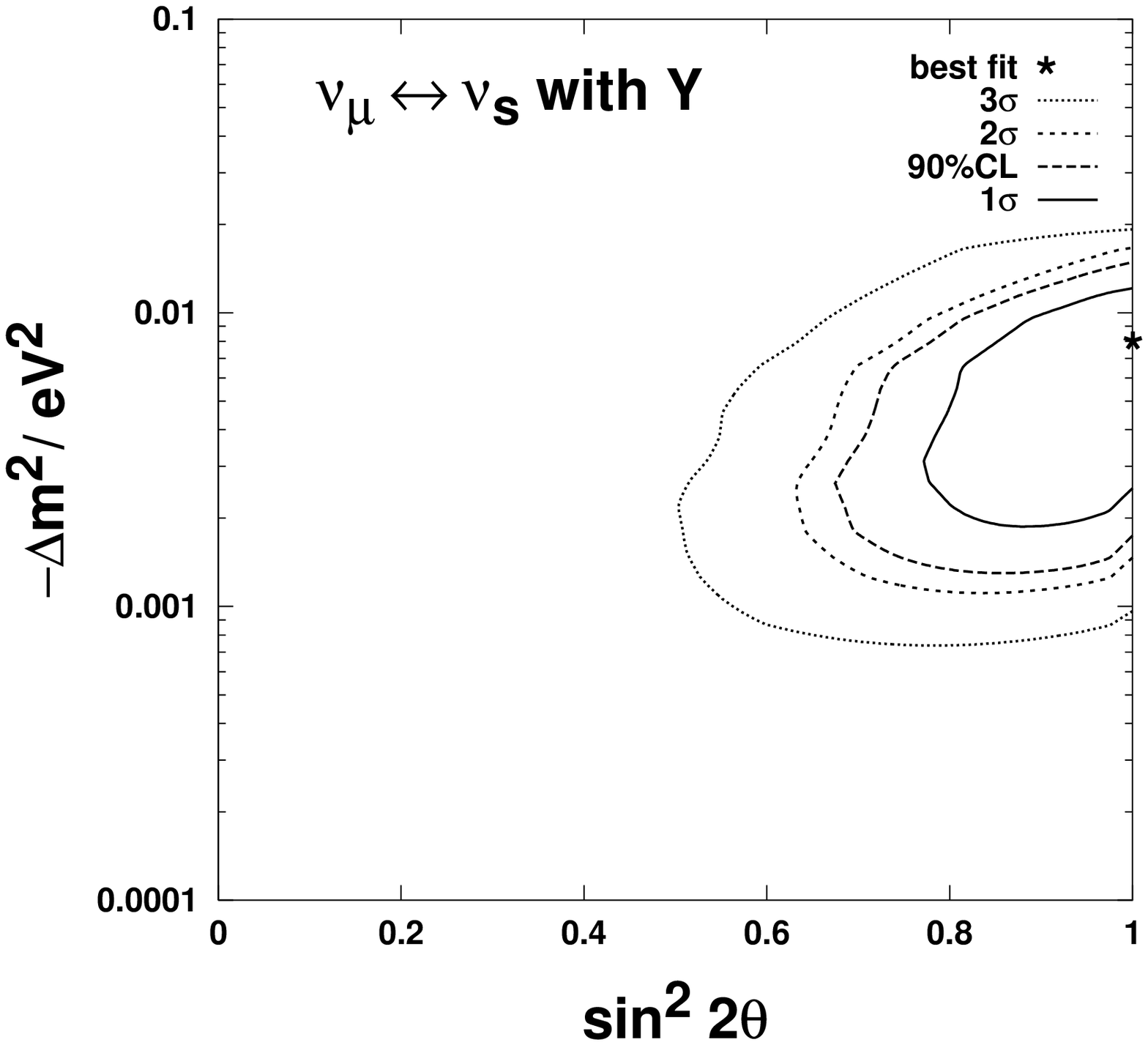}\hfil}
\caption{\label{fig:fig8}The same as Figure 7 but with $R$ data
excluded from the fit.}
\end{figure}

The results of the $\chi^2$ fits are displayed in Figs.1--8.
In Figs. 1 and 2, $\chi^2$ is plotted against $\Delta m^2$.
For $\nu_\mu\leftrightarrow\nu_\tau$, $\chi^2$
does not experience a deep
minimum at the best fit point with respect to $\Delta m^2$
particularly when the $R$'s are excluded from the fit.
In general, for geometrical reasons,
atmospheric neutrino analysis does not constrain
$\Delta m^2$ very precisely.
Note that the the situation is
slightly different in case of $\nu_\mu\leftrightarrow\nu_s$.
Figure 3 shows the allowed region of
$(\sin^2 2\theta,\ \Delta m^2)$
at various confidence levels for the $\nu_\mu \leftrightarrow\nu_\tau$
scenario.  Maximal mixing provides the best fit, and $\Delta m^2$
values in the $10^{-3}$ to $10^{-2}$ eV$^2$ range are favored.  Note
that the confidence levels are defined in the usual way by
\begin{eqnarray}
\chi^2 = \chi^2_{\rm min} + \Delta \chi^2\nonumber
\end{eqnarray}
where $\Delta \chi^2 = 2.3, 4.6, 6.2, 11.8$ for the 
$1\sigma$, $90\%$ C.L., $2\sigma$ and 
$3\sigma$ allowed region respectively.
Our $\chi^2_{\rm min}$ for $\nu_\mu \leftrightarrow\nu_\tau$ 
oscillations is $\chi^2_{\rm min} = 4.5$
for $4$ degrees of freedom. This is quite a good fit to the data (allowed
at the $35\%$ level).
In Figure 4 we show the allowed region considering just
the asymmetries instead of using both the asymmetries and 
the $R$ ratios.
Note that in this case there are 4 data points
and 2 free parameters which gives 2 degrees
of freedom.

Figures 5--8 show the corresponding results for the $\nu_\mu
\leftrightarrow\nu_s$ scenario. If $\Delta m^2>0$, smaller values of
$\Delta m^2$ are disfavored because the matter effect moves both $R$
and $Y$ away from the measured values, but if $\Delta m^2<0$, then
smaller values of $\Delta m^2$ and $\sin^22\theta$ are permitted at
the 90\% confidence level.  The value of $\chi^2_{\rm min}$ for the
$\nu_\mu \leftrightarrow\nu_s$ scenario is $\chi^2_{\rm min} = 5.1$ for
$4$ degrees of freedom. This is similar to $\nu_\mu - \nu_\tau$ case
and also represents quite a good fit (which is allowed at $28\%$).

To summarize, both the solutions $\nu_\mu \leftrightarrow\nu_\tau$ and
$\nu_\mu \leftrightarrow\nu_s$ provide a good fit to the contained
events of the Superkamiokande atmospheric neutrino data.

\section{Other analyses}

There have been several proposals to distinguish the
$\nu_\mu\leftrightarrow\nu_\tau$ and $\nu_\mu\leftrightarrow\nu_s$
oscillations.
Matter effects in $\nu_\mu\leftrightarrow\nu_s$ oscillations in upward
going muon data were first analyzed by Akhmedov, Lipari and Lusignoli
\cite{all} and more recently by Lipari and Lusignoli \cite{ll}.  It
has been pointed out by Liu, Smirnov \cite{ls} and
Liu, Mikheyev, Smirnov \cite{lms} that
signatures due to parametric enhancement in
$\nu_\mu\leftrightarrow\nu_s$ oscillations may be seen in upward going
muon data.
Vissani and Smirnov \cite{vs} proposed to look at the ratio
($\pi^0$-events)/(two ring events).  Learned, Pakvasa and Stone
\cite{lps} suggested that the up-down asymmetry (upward going
$\pi^0$-events)/(downward going $\pi^0$-events) can tell a difference.
Hall and Murayama \cite{hm} proposed a similar technique to use the
up-down asymmetry in the multi-ring events.  Kajita \cite{kajita}
mentioned the ratio ($\pi^0$-events)/(e-like events) which should in
principle enable us to distinguish.
All these analyses seem to be still inconclusive and we need more
statistics and accurate knowledge on nuclear cross sections to draw a
conclusion.


\section{Conclusions}

We have demonstrated that matter effects in the Earth have a
significant role to play in comparing and contrasting the
$\nu_\mu\leftrightarrow\nu_\tau$ and $\nu_\mu\leftrightarrow\nu_s$
solutions to the atmospheric neutrino anomaly with Superkamiokande
data.  So far both solutions provide a good fit to the data and we
need more statistics to be conclusive.  
We hope that non-accelerator
experiments such as Superkamiokande will distinguish them before
future long baseline experiments with emulsion techniques
\cite{emulsion} give direct evidence.


\section*{Acknowledgments}

The author would like to thank R. Foot and R.R. Volkas for
collaboration, discussions and comments on this manuscript.  He also
would like to thank T. Kajita and E. Kearns for useful communications.
This research was supported in part by a Grant-in-Aid for Scientific
Research of the Ministry of Education, Science and Culture,
\#09045036, \#10140221, \#10640280.



\begin{thebibliography}{99}
\bibitem{lps}
J.G. Learned, S. Pakvasa and J.L. Stone, hep-ph/9805343.
The author has learned from E. Kearns that this technique
has been thought about for some time by the members of
the Superkamiokande collaboration.
\end{thebibliography}
\end{document}